\def \ba {\begin{array}}
\def \ea {\end{array}}

\def \bea {\begin{eqnarray}}
\def \eea {\end{eqnarray}}

\def \be {\begin{equation}}
\def \ee {\end{equation}}

\def\nn{\nonumber}

\def\l[{\left[}
\def\r]{\right]}

\documentclass[12pt,a4]{article}
\usepackage{amssymb,amsfonts}
\begin{document}

\begin{center} 
{\bf A first analysis regarding matter-dynamical diffeomorphism coupling}
\footnote{Work partially
supported by the DGICYT.}
\end{center}
\bigskip
\bigskip
\centerline{ {\it V. Aldaya\footnote{E-mail: valdaya@iaa.es} 
and J.L. Jaramillo\footnote{E-mail: jarama@iaa.es}} }
\bigskip

\begin{itemize}
\item {Instituto de Astrof\'{\i}sica de Andaluc\'{\i}a (CSIC), Apartado Postal
 3004, 18080 Granada, Spain.}
\item  {Instituto Carlos I de F\'\i sica Te\'orica y Computacional, Facultad
de Ciencias, Universidad de Granada, Campus de Fuentenueva, 
Granada 18002, Spain.} 
\end{itemize}

\bigskip
\begin{center}
{\bf Abstract}
\end{center}
\small

\begin{list}{}{\setlength{\leftmargin}{3pc}\setlength{\rightmargin}{3pc}}
\item 
A first attempt at adding {\it matter} degrees of freedom to the 
two-dimensional ``vacuum'' gravity model presented in \cite{qg1199} is 
analyzed 
in this paper. Just as in the previous pure gravity case, quantum 
diffeomorphism operators (constructed from a Virasoro algebra) possess a 
dynamical content; their gauge nature is recovered only after the classical 
limit. Emphasis is placed on the new physical modes modelled on a 
$SU(1,1)$-Kac-Moody algebra. The non-trivial coupling to ``gravity'' is a 
consequence of the natural semi-direct structure of the entire extended 
algebra. A representation associated with the discrete series of the rigid 
$SU(1,1)$ is revisited in the light of previously neglected crucial global 
features which imply the appearance of an $SU(1,1)$-Kac-Moody fusion rule, 
determining the rather entangled
quantum structure of the physical system. In the classical limit, an action 
which explicitly couples gravity and matter modes governs the dynamics.  
\end{list}

\normalsize

\vskip 1cm

\section{Introduction}
A model for dealing with the potential dynamical content of the operators
associated with diffeomorphisms
in the quantum regime was recently presented in \cite{qg1199}. Accepting
the (centrally-extended) abstract Virasoro group as the only physical input of
the theory, a spacetime notion inside the group was found for only
a critical combination of the central extension parameters. The resulting
physical model presented an ensemble of coexisting spacetimes, mixed 
at the quantum level by the action of diffeomorphisms, while a correction
to Polyakov's two-dimensional gravity action was found in the semi-classical 
limit.

The main goal of this paper is the study of the coupling 
of new dynamical degrees of freedom, eventually interpretable as {\it matter},
to the previous model. Symmetry being our unique physical guide, we can 
resort only to algebraic structures in order to gain intuition regarding the
manner of inserting the new modes. In particular, we are interested in a Lie 
algebra
structure containing the Virasoro algebra as a subalgebra, in order to 
incorporate the {\it pure gravity} model, and coupling the diffeomorphism
algebra to the new modes in a non-trivial way. It is by no means obvious
that the enlarging the original algebra preserves the {\it critical} way in 
which spacetime emerged; that is, as associated with a critical value of the 
conformal charge.

The abstract approach followed here has the advantage of a clear and 
non-ambiguous definition of the mathematical structure of the physical system,
but poses the serious problem 
of its physical interpretation. We have no other alternative than
offering an {\it a posteriori} interpretation of the model depending of the 
concrete realization of the system. In some sense, any fundamental theory 
(as opposed to an effective one) must confront and solve this pitfall.

The simplest Lie algebra that fulfils the previous requirements is an
affine Kac-Moody Lie algebra under the semi-direct action of the Virasoro 
group, and this will be our choice in the present paper. More 
specifically,
and motivated by Julia and Nicolai's analysis \cite{Nicolai} of 
two-dimensional gravity, where matter fields live on the quotient space of a 
finite-dimensional non-compact group, we shall choose a non-compact
Kac-Moody group to implement the new degrees of freedom. Again
on behalf of simplicity in this first approach to the problem, our choice
will be the most manageable one; that is, we study $SU(1,1)$-Kac-Moody.
This symmetry has also recently been considered for constructing coset
models for black holes \cite{Witten}.
Thus, the Lie algebra ${\cal G}$ defining our physical system is: 
\bea
[\hat{Z}_n,\hat{Z^*}_m]&=&-2i \hat{\Phi}_{n+m}-2 (\alpha n+
\frac{K}{2})\delta_{n+m}\hat{I} \nn \\
\left[ \hat{\Phi}_n,\hat{Z}_m \right] &=& i \hat{Z}_{n+m}\;
,\;
\left[\hat{\Phi}_n,\hat{Z^*}_m\right]=-i \hat{Z^*}_{n+m} \nn \\
\left[\hat{\Phi}_n,\hat{\Phi}_m\right]&=&\alpha n \delta_{n+m}\hat{I} 
\nn \\
\left[\hat{L}_n,\hat{Z}_m\right]&=&-m \hat{Z}_{n+m}\; , \; 
\left[\hat{L}_n,\hat{Z^*}_m\right]=-m \hat{Z^*}_{n+m}
\label{algebra} \\
\left[\hat{L}_n,\hat{\Phi}_m\right]&=&-m \hat{\Phi}_{n+m}+im
\frac{K}{2}\delta_{n+m}\hat{I} \nn \\
\left[\hat{L}_n,\hat{L}_m\right]&=&(n-m)\hat{L}_{n+m}+\frac{1}{12}(cn^3-c'n)
\delta_{n+m}\hat{I} \nn \\
\left[\hat{I},...\right]&=&0 \nn
\eea
The mathematical formalism to be used to derive the physical system
from the previous algebra is the Group Approach to
Quantization (GAQ). A non-technical presentation of it can be found in 
\cite{qg1199} (see references therein for deeper and more rigorous 
explanations). Here we just
outline the basic features, which will be treated with more detail as they
appear in the main text.

Our first step will be to consider the possible central extensions 
and pseudo-extensions\footnote{Redefinitions of certain generators with 
non-trivial dynamical consequences. They are crucial for semi-simple 
groups, where true cohomology is trivial.} of the 
algebra, which are determined by its cohomology
and pseudo-cohomology, respectively. This allows a distinction (if no anomalous
reduction is present at the representation level) between the dynamical (or
basic) degrees of freedom and the kinematical ones. The first set define the
physical phase space of the system, whereas the second one induces evolution 
flows along the former. In (\ref{algebra}) the cohomology has already been 
taken into account.

After cohomology is analysed, the next crucial point is to {\it exponentiate} 
the Lie algebra ${\cal G}$ to a Lie group $G$. Dynamics are contained in the 
group rather than
in the algebra, and global questions, which become visible at the group level 
but
not at the algebra one, prove to be critical in some points of our analysis. In
this sense, we shall see in {\bf subsection 3.2} how imposing of 
globallity
on the wave functions brings about the presence of restrictions in the 
representation that guarantee the consistency (unitarity) of the theory. 
Once we have a centrally-extended group law, we can compute the left- and
right-invariant vector fields as well as the quantization one-form $\Theta$ 
(that is, the component of the canonical left-invariant one-form which is 
dual to the central vector field $\hat{I}$). The kinematical vector fields, 
those
comprising the kernel of the Lie-algebra two-cocycle, can be characterised 
as spanning the left-invariant characteristic subalgebra ${\cal G}_\Theta=
Ker(\Theta)\cap Ker(d\Theta)$. With these elements at hand, one can develop a 
semi-classical formalism, or a directly quantum one. 

The main element that defines the semi-classical formalism 
is the classical phase space ${\cal M}$, obtained by taking the quotient of 
the group by the equations of
motion as well as by the $U(1)$ central extension subgroup 
(${\cal M}=G/({\cal G}_
\Theta\otimes U(1)$). It can be parametrized in terms of the basic Noether 
invariants, $i_{X^R_i}\Theta$, where the respective ${X^L_i}$ are not present 
in the characteristic subalgebra. The symplectic form is defined by $d\Theta$, 
which properly falls down to the quotient. Finally, an action $S$ can be 
derived by integrating $\Theta$ over trajectories on the group: $S=\int 
\Theta$. This formalism is the one used at the classical level.

The quantum theory is obtained after reducing the regular representation, 
defined by the action of the right-invariant vector fields over the 
complex $U(1)$-functions with support on the group, by means of a polarization
(see
{\bf subsection 3.1}) constructed by using left-invariant vector fields. These 
latter
commute with the right-invariant ones, thus respecting the group action. The
resulting representation must be unitary in order to introduce a standard 
quantum probability notion.

The organization of the paper parallels that of \cite{qg1199} and is as 
follows. In {\bf section 2} we present the elements of the semi-classical 
formalism, emphasizing those appearing in the classical 
limit. {\bf Section 3} deals with the quantum theory and is divided into three
subsections which deal with reducibility, unitarity, and a possible 
contraction 
phenomenon in the theory, respectively. With the previous elements in mind,
{\bf section 4} tries to give a physical interpretation to the mathematical 
objects introduced in the preceding sections. Finally, several comments and
reflections are presented in {\bf section 5}.

Before we start, let us insist on the fact that the reasoning and results in 
the following sections must be contemplated in the spirit and context of
\cite{qg1199}.

\section{Semi-classical formalism}
As explained in the previous section, our first step in the construction of 
the model is the {\it exponentiation} to a group law from the Lie algebra 
(\ref{algebra}), which defines the physical system, since physics are encoded 
in the (global) Lie group structure.

To accomplish this goal, we use the $SU(1,1)$ local group law:
\bea
z''&=&z\eta'^{-2}+\kappa z'+\frac{2 z'}{1+\kappa'}[{z^*}' z\eta'^{-2}+z^*z'
\eta^2]
\nn\\
{z^*}''&=&z^*\eta'^{2}+\kappa {z^*}'+\frac{2 {z^*}'}{1+\kappa'}[z'z^*\eta'^2+
z{z^*}'\eta^{-2}] \\
\eta''&=&\sqrt{\frac{2}{1+\kappa''}}\left[\sqrt{\frac{1+\kappa}{2}}
\sqrt{\frac{1+\kappa'}{2}}\eta\eta'+\sqrt{\frac{2}{1+\kappa}}
\sqrt{\frac{2}{1+\kappa'}}{z^*}'z{\eta^*}'\eta\right]\;, \nn
\eea
where
\bea 
\kappa&=&\sqrt{1+2zz^*}\;\;;\;\;\kappa''=\kappa\kappa'+z{z^*}'\eta'^{-2}+
z^*z'\eta'^2 \nn \\
\eta&=&e^{i\phi} \nn
\eea
The law for the Virasoro subgroup can be found in \cite{qg1199} and references
therein. Expanding the elements of $SU(1,1)$ in a formal Laurent series to 
conform to the loop group law \footnote{We expand up to the third order in
the parameters, which is sufficient for our purposes.}, and 
using the techniques developed in \cite{KMpepe} to construct the cocycles
for the central extensions as well as the (semi-direct) action of the Virasoro
subgroup on the $SU(1,1)$-Kac-Moody subgroup, we find the following group law:
\bea
{z^n}''&=&z^n+A^n_k(l){z^k}'-iA^m_k(l)z^{n-m}\phi^k+A^m_k(l)A^p_l(l)
(z^{n-m-p}{z^k}'{{z^*}^l}'+{z^*}^{n-m-p}{z^k}'{z^l}' \nn \\
&-&\frac{1}{2}z^{n-m-p}{\phi^k}'{\phi^l}')+2A^p_l(l)z^{n-m-p}{z^*}^m{z^l}'
+... \nn \\
{{z^*}^n}''&=&{z^*}^n+A^n_k(l){{z^*}^k}'+iA^m_k(l){z^*}^{n-m}\phi^k+
A^m_k(l)A^p_l(l)({z^*}^{n-m-p}{{z^*}^k}'{z^l}'+z^{n-m-p}{{z^*}^k}'{{z^*}^l}' 
\nn \\
&-&\frac{1}{2}{z^*}^{n-m-p}
{\phi^k}'{\phi^l}')+2A^p_l(l){z^*}^{n-m-p}z^m{{z^*}^l}'+... \nn \\
{\phi^n}''&=&\phi^n+A^n_k(l)\phi^k-iA^m_k(l)(z^{n-m}{{z^*}^k}'
-{z^*}^{n-m}{z^k}')-A^m_k(l)A^p_l(l)(z^{n-m-p}{{z^*}^k}' \\
&+&{z^*}^{n-m-p}{z^k}')\phi^l+... \nn \\
l^{\prime\prime m}&=&l^{m}+l'^{m}+ipl'^{p}l^{m-p}+
\frac{(ip)^2}{2!}l'^{p}l^{n}l^{m-n-p}+...+
\sum\limits_{ n^1+...+n^j+p=m}\frac{(ip)^j}{j!}l'^{p}
l^{n^1}...l^{n^j}+...\nn \\
\varphi''&=&\varphi+\varphi'+\frac{K}{2}\xi_{cobKM}+\alpha \xi_{KM}+
\frac{c}{24}\xi_{Vir}-\frac{c'}{24}\xi_{cobVir} \nn
\eea
where
\bea
A^k_n(l)&=&\delta^k_n+(in)l^{k-n}+\sum\limits_{r=2}\sum\limits_{n+m_1+...
+m_r=k}\frac{{(in)}^r}{r!}l^{m_1}...l^{m_r} \nn \\
\xi_{KM}&=&\frac{-1}{2}[inA^{-n}_k(l)(2{z^n}{{z^*}^k}'+2{{z^*}^n}{z^k}'-
\phi^n{\phi^k}')- iA^m_k(l)A^{-n-m}_k(l)i(n-m)(z^n{{z^*}^k}'{\phi^l}' 
\nn\\
&-&{z^*}^n{z^k}'{\phi^l}')-iA^{-n-m}_k(l)in({z^*}^m\phi^n{z^k}'
-z^m\phi^n{{z^*}^k}')]+... \label{A} \\
\xi_{Vir}&=&-[(-i)(-n)n^2{l^{-n}}'l^n+\frac{(-i)^2}{2!}n_1n_2(n_1+n_2)^2
{l^{n_1}}'{l^{n_2}}'l^{-n_1-n_2} \nn \\
&-&\frac{i^2}{2!}(n_1+n_2)^2
(n_1^2+n_2^2+n_1n_2){l^{-n_1-n_2}}'l^{n_1}l^{n_2}+...] \nn \\
\xi_{CobKM}&=&{\phi^0}''-{\phi^0}'-\phi^0 \nn \\
\xi_{CobVir}&=&{l^0}''-{l^0}'-l^0 \nn. 
\eea
With the explicit group law at our disposal, the left- and right-invariant 
vector fields can be systematically computed, obtaining:
\bea
\tilde{X}_{z^r}^L&=&\frac{\partial}{\partial z^r} -i\phi^{n-r}
\frac{\partial}{\partial z^n}+z^{n-p-r}{z^*}^p\frac{\partial}{\partial z^n}-
\frac{1}{2}\phi^{n-p-r}\phi^p\frac{\partial}{\partial z^n}+{z^*}^{n-p-r}
{z^*}^p\frac{\partial}{\partial {z^*}^n}-i{z^*}^{n-r}\frac{\partial}
{\partial \phi^n}\nn \\
&-&{z^*}^{n-p-r}\phi^p\frac{\partial}{\partial \phi^n}+...+
[-i(\frac{K}{2}+\alpha r){z^*}^{-r}-(\frac{K}{2}+\frac{\alpha}{2}(r-m))
{z^*}^m\phi^{-m-r}+...]\frac{\partial}{\partial \varphi} \nn \\
\tilde{X}_{{z^*}^r}^L&=&\frac{\partial}{\partial {z^*}^r} +i\phi^{n-r}
\frac{\partial}{\partial {z^*}^n}+{z^*}^{n-p-r}z^p\frac{\partial}
{\partial {z^*}^n}-
\frac{1}{2}\phi^{n-p-r}\phi^p\frac{\partial}{\partial {z^*}^n}+
z^{n-p-r}z^p\frac{\partial}{\partial {z^*}^n}+iz^{n-r}\frac{\partial}
{\partial \phi^n}\nn \\
&-&z^{n-p-r}\phi^p\frac{\partial}{\partial \phi^n}+...+
[i(\frac{K}{2}-\alpha r)z^{-r}-(\frac{K}{2}-\frac{\alpha}{2}(r-m))
z^m\phi^{-m-r}+...]\frac{\partial}{\partial \varphi}  \\
\tilde{X}_{\phi^r}^L&=&\frac{\partial}{\partial \phi^r}+
(\frac{\alpha}{2}ir\phi^{-r}+...)\frac{\partial}{\partial \varphi} \nn \\
\tilde{X}_{l^r}^L&=&\tilde {X}_{l^r}^{L\;Vir}+i(n-r)z^{n-r}
\frac{\partial}{\partial z^n}+i(n-r){z^*}^{n-r}\frac{\partial}{\partial 
{z^*}^n}+i(n-r)\phi^{n-r}\frac{\partial}{\partial \phi^n} \nn \\
\Xi&=&\tilde{X}^L_\varphi=\frac{\partial}{\partial \varphi} \nn
\eea 
for the left-invariant vector fields, and
\bea
\tilde{X}_{z^r}^R&=&A^n_r(l)\frac{\partial}{\partial z^n}+2A^p_r(l)z^{n-m-p}
{z^*}^m\frac{\partial}{\partial z^n}+iA^m_r(l){z^*}^{n-m}\frac{\partial}
{\partial \phi^n}+... \nn \\
&+&[i(\frac{K}{2}-\alpha n)A^{-n}_r(l){z^*}^n
-\frac{\alpha}{2}nA^{-n-m}_r(l){z^*}^m\phi^n+...]\frac{\partial}
{\partial \varphi} \nn \\
\tilde{X}_{{z^*}^r}^R&=&A^n_r(l)\frac{\partial}{\partial {z^*}^n}+
2A^p_r(l){z^*}^{n-m-p}
z^m\frac{\partial}{\partial {z^*}^n}-iA^m_r(l)z^{n-m}\frac{\partial}
{\partial \phi^n}+... \nn \\
&+&[-i(\frac{K}{2}+\alpha n)A^{-n}_r(l)z^n
+\frac{\alpha}{2}nA^{-n-m}_r(l)z^m\phi^n+...]\frac{\partial}
{\partial \varphi}  \\
\tilde{X}_{\phi^r}^R&=&A^n_r(l)\frac{\partial}{\partial \phi^n}-
iA^m_r(l)z^{n-m}\frac{\partial}{\partial z^n}+iA^m_r(l){z^*}^{n-m}\frac
{\partial}{\partial {z^*}^n}+... +(\frac{\alpha}{2}nA^{-n}_r(l)\phi^n+...)\frac
{\partial}{\partial \varphi} \nn \\
\tilde{X}_{l^r}^R&=&\tilde{X}_{l^r}^{R\;Vir}=\frac{\partial}{\partial l}+
irl^{m-r}\frac{\partial}{\partial l^m}+ [\frac{ic}{24} r^3 l^{-r}-\frac{c}{24}
r^2\sum\limits_{n_1+n_2=-r}(n_1^2+n_2^2+n_1n_2)l^{n_1}l^{n_2}\nn \\
&-&\frac{ic'}{24}
rl^{-r}+\frac{c'}{24} \sum\limits_{n_1+n_2=-r}\frac{r^2}{2}l^{n_1}l^{n_2}+...]
\frac{\partial}{\partial \varphi} \nn \\
\Xi&=&\tilde{X}^R_\varphi=\frac{\partial}{\partial \varphi}\;\;\;\;, \nn
\eea
where $\tilde {X}_{l^r}^{L Vir}$ and $\tilde {X}_{l^r}^{R\;Vir}$ are the 
expressions that can be found in \cite{qg1199}.

In order to obtain the quantization one-form $\Theta$, that is, the vertical 
component of the canonical left-invariant one-form on the group, we impose 
duality on the left-invariant vector fields ($\Theta(\tilde{X}_{z^r}^L)=
\Theta(\tilde{X}_{{z^*}^r}^L)=
\Theta(\tilde{X}_{\phi^r}^L)=\Theta(\tilde{X}_{l^r}^L)=0$ and 
$\Theta(\Xi)=1$). The resulting expression is:
\bea
\Theta&=&\Theta^{KM}+\Theta^{Vir}+\Theta^{Int}+d\varphi \nn \\
\Theta^{KM}&=&\frac{-i\alpha}{2}r\phi^{-r}d\phi^r+[i(\frac{K}{2}+\alpha r)
{z^*}^{-r}+\alpha(n+r){z^*}^n\phi^{-n-r}+...]dz^r+[i(\frac{-K}{2}+\alpha r)
z^{-r} \nn \\
&-&\alpha(n+r)z^n\phi^{n-r}+...]d{z^*}^r  \nn \\
\Theta^{Vir}&=& \frac{i}{24}(cn^2-c')nl^{-n}dl^n+  \label{teta}\\
&+&\sum_{\stackrel{k=2}{n_1+...n_k=-n}}
\frac{(-i)^k}{24}[cn_1^2-c'+ 
cn^2\sum_{m=2}^k\frac{1}{m!}]n_1...n_kl^{n_1}...l^{n_k}dl^n \nn \\
\Theta^{Int}&=&\sum_{\stackrel {j=0}{n_1+...+n_j+k=-r}}(-i)^j f^k(z^r,{z^*}^r,
\phi^r)n_1...n_jdl^r \;\;\;\;, \nn 
\eea
where
\bea
f^k(z^r,{z^*}^r,\phi^r)&=&-i\alpha(n+k)nz^{-n+k}{z^*}^n-\frac{iK}{2}(n+k)
z^{n+k}{z^*}^{-n}-i\alpha(n+k)n{z^*}^{-n+k}z^n \nn \\
&+&\frac{iK}{2}(n+k){z^*}^{n+k}
z^{-n}+\frac{i\alpha}{2}n(n+k)\phi^{-n}\phi^{n+k}+\alpha n m z^n\phi^m
{z^*}^{k-n-m}  \\
&-&\alpha n m {z^*}^n\phi^mz^{k-n-m}+...\;\;\;\;. \nn
\eea
For both the classical and the quantum theory, the identification of the 
characteristic subalgebra ${\cal G}_{\Theta}$ ($=Ker\Theta\cap Kerd\Theta$), 
is a 
crucial point. The central extension structure simplifies this search, reducing
it to the determination of the kernel of the Lie-algebra cocycle. Considering
the commutation relationships (\ref{algebra}) we notice that the composition
of ${\cal G}_{\Theta}$ depends of the actual values of the central extensions
$c$ and $\alpha$, as well as pseudo-extensions $c'$ and $K$, giving rise to a 
wide range of possibilities. We shall study the specific choice of the 
extension 
parameters that fits our physical purposes. Following \cite{qg1199}, we want to
find a $sl(2,\mathbb R)$ (Virasoro-)subalgebra inside the characteristic 
subalgebra in order to construct a spacetime notion. This can be achieved if 
we impose $c=c'-3\frac{K^2}{\alpha}$ \footnote{We shall see in the next 
section that quantum theory imposes a correction to this relationship.}. In
fact, the linear combination,
$\tilde{\bar{X}}_{l^{i}}^L=\tilde{X}_{l^{i}}^L
+\frac{K}{2\alpha}\tilde{X}_{\phi^i}^L$ $(i\in\{-1,0,1\})$, then enters
${\cal G}_\Theta$. 
Furthermore, we require 
$\frac{K}{2\alpha}\not\in{\mathbb Z}$ in 
order not to lose dynamical modes in the physical (matter) fields.
Then we have:
\bea 
{\cal G}_{\Theta}=\langle \tilde{X}_{\phi^0}^L,\tilde{\bar{X}}_{l^{-1}}^L,
\tilde{\bar{X}}_{l^0}^L,\tilde{\bar{X}}_{l^{1}}^L\rangle \;\;\;\;.
\eea

The linear combination giving rise to the $sl(2,\mathbb R)$ inside 
${\cal G}_\Theta$ suggests the possibility of generalizing it for the rest of 
the Virasoro modes, closing again a Virasoro subalgebra 
($\overline{Vir}$):
\bea
\tilde{X}_{l^{n}}^L\rightarrow\tilde{\bar{X}}_{l^{n}}^L=\tilde{X}_{l^{n}}^L+
\frac{K}{2\alpha}
\tilde{X}_{\phi^n}^L\;\;\;,\;\;\; \forall n \;\;.
\eea
In that case, the pure Kac-Moody commutation relations remain the same, but 
the Virasoro ones take the form:
\bea
\left[\tilde{\bar{X}}^L_{l^n},\tilde{X}^L_{z_m}\right]&=&-i(m+
\frac{K}{2\alpha}) 
\tilde{X}^L_{z_{n+m}}\; , \; 
\left[\tilde{\bar{X}}^L_{l^n},\tilde{X}^L_{z^*_m}\right]=-i(m-
\frac{K}{2\alpha}) 
\tilde{X}^L_{z^*_{n+m}} \nn \\
\left[\tilde{\bar{X}}^L_{l^n},\tilde{X}^L_{\phi_m}\right]&=&-im \tilde{X}^L_
{\phi_{n+m}} \\
\left[\tilde{\bar{X}}^L_{l^n},\tilde{\bar{X}}^L_{l^m}\right]&=&i(n-m)\tilde
{\bar{X}}^L_{l^{n+m}}+
\frac{i}{12}(cn^3-(c'-3\frac{K^2}
{\alpha})n)\delta_{n+m}\Xi \;\;\;\;. \nn  
\eea
This basis will be more suited for the polarization analysis in the next 
section.

The classical equations of motion for the model consist of the 
dynamical 
system defined in terms of the vector fields in ${\cal G}_{\Theta}$, which 
dictate the evolution of the parameters in the group. Looking at the 
explicit form of the left-invariant vector fields, we observe that the 
group parameters are constant under $\tilde{X}^L_{\phi^i}$, so we can ignore
them in the equations of motion. Thus the evolution is parametrized solely by
the {\it space-time} $SL(2,\mathbb R)$ subgroup: 
\be
\frac {\partial g_i^n} {\partial \tilde{\lambda}_0}=(\tilde{X}_{l^0}^{L})
g_i^n
\hbox{\ \ , \ \ }\frac {\partial g_i^n} {\partial \tilde{\lambda}_1}=
(\tilde{X}_{l^1}^{L})g_i^n\hbox{\ \ , \ \ }\frac {\partial g_i^n} {\partial
\tilde{\lambda}_{-1}}=(\tilde{X}_{l^{-1}}^{L})g_i^n\;\;\;,\;\;\; 
i\in\{1,2,3,4\}\; ,
\ee
where $g_1^n=z^n,g_2^n={z^*}^n,g_3^n=\phi^n,g_4^n=l^n$; while 
$\tilde{\lambda}_0,\tilde{\lambda}_1$ and $\tilde{\lambda}_{-1}$ are the 
parameters of the vector fields $\tilde{X}_{l^0}^{L},\tilde{X}_{l^1}^{L}$ and
$\tilde{X}_{l^{-1}}^{L}$, respectively.

The following explicit equations of 
motion are {\it exact}, in spite of the development of the Kac-Moody subgroup
up to the third order.
\bea
\frac {\partial g_i^m} {\partial \tilde{\lambda}_0}&=&im g_i^{m}  
\hbox{ \ , \ }
\frac {\partial g_i^m} {\partial \tilde{\lambda}_1}=i(m-1)g_i^{m-1} 
\hbox{ \ , \ }
\frac {\partial g_i^m} {\partial\tilde{\lambda}_{-1}}=i(m+1)g_i^{m+1}
\hbox{ , }i\in\{1,2,3\}\nn \\
\frac {\partial l^m} {\partial \tilde{\lambda}_0}&=&im l^{m}\hbox{\ \ \ for
 \ \ \ } 
 m\neq0 
\hbox {\ \ \ , \ \ \ }
\frac {\partial l^0} {\partial \tilde{\lambda}_0}=1 \nn \\
\frac {\partial l^m} {\partial \tilde{\lambda}_1}&=&
i(m-1)l^{m-1}\hbox{\ \ \ for \ \ \ } 
 m\neq1 
\hbox {\ \ \ , \ \ \ }
\frac {\partial l^1} {\partial\tilde{\lambda}_1}=1 \label{motion}\\
\frac {\partial l^m} {\partial\tilde{\lambda}_{-1}}&=&
i(m+1)l^{m+1}  \hbox{\ \ \ for \ \ \ } 
 m\neq-1 
\hbox {\ \ \ , \ \ \ }
\frac {\partial l^{-1}} {\partial\tilde{\lambda}_{-1}}=1 \hbox{ .} \nn 
\eea
These equations have the same form as those found in \cite{qg1199} (in fact, 
they
are exactly the same for the $l^n$), so they can be solved exactly and present
the structure: 
\bea
g_i^n(\lambda_{-1},\lambda_0,\lambda_1)&=&g_i^n(\lambda_{-1},\lambda_1)
e^{in\lambda_0} \hbox{ \ \ , \ \ }i\in\{1,2,3\} \\
l^n(\lambda_{-1},\lambda_0,\lambda_1)&=&l^n(\lambda_{-1},\lambda_1)
e^{in\lambda_0} \hbox{ , }n\neq 0 \hbox{ \ ; \ } l^{0}=\lambda_0 \;\;, \nn
\eea
where $\tilde{\lambda}_0= \lambda_0$, $\tilde{\lambda}_1=\lambda_1
e^{i\lambda_0}$, $\tilde{\lambda}_{-1}=\lambda_{-1}e^{-i\lambda_0}$.

The symplectic manifold characterizing the classical physical system is 
obtained by taking the quotient of the (non-extended) group by the  
equations of motion. The symplectic form is defined by $d\Theta$, which 
passes to the quotient. This phase space can be parametrized by the Noether
invariants of the group parameters whose left-invariant vector fields are not 
present in ${\cal G}_\Theta$.
This is the reason for referring to these modes as dynamical (or basic) 
degrees 
of freedom. We give the explicit expressions for the Noether invariants up
to the second order:
\bea
Z_r&=&i_{\tilde{X}_{z^r}^R}\Theta=2i(\frac{K}{2}+\alpha r){z^*}^{-r}+2\alpha
(m+r){z^*}^m\phi^{-m-r}-2r(\frac{K}{2}+\alpha n)l^{n-r}z^{-n}+... \nn \\
{Z^*}_r&=&i_{\tilde{X}_{{z^*}^r}^R}\Theta=-2i(\frac{K}{2}-\alpha r)z^{-r}-
2\alpha(m+r)z^m\phi^{-m-r}-2r(\frac{K}{2}-\alpha n)l^{n-r}{z^*}^{-n}+...  \\
\Phi_r&=&i_{\tilde{X}_{\phi^r}}^R \Theta=-i\alpha r\phi^{-r}+
z^{m-r}{z^*}^{-m}[(\frac{K}{2}+\alpha m)-(\frac{-K}{2}+\alpha(r-m))]+
\alpha nr\phi^{-n}l^{n-r}+...\nn \\
L_r&=&i_{\tilde{X}_{l^r}}^R\Theta=L_r^{Vir}+\frac{i\alpha}{2}n(-n-r)[\phi^n
\phi^{-n-r}-4z^{-n-r}{z^*}^n]-\frac{Ki}{2}(2n-r)z^{n-r}{z^*}^{-n}
+... \;\;,\nn			
\eea
where, again, the superscript $Vir$ denotes the object that can be found in 
\cite{qg1199}.

The configuration-like description of the system, better suited for a 
Lagrangian formalism, can be obtained after defining the fields:
\bea
F_z(\lambda_{-1},\lambda_0,\lambda_1)&\equiv&\sum\limits_{n}z^n(\lambda_{-1},
\lambda_0,\lambda_1)= \sum\limits_{n}z^n(\lambda_{-1},\lambda_1)
e^{in\lambda_0} \nn\\
F_{z^*}(\lambda_{-1},\lambda_0,\lambda_1)&\equiv&\sum\limits_{n}{z^*}^n
(\lambda_{-1},\lambda_0,\lambda_1)=\sum\limits_{n}{z^*}^n(\lambda_{-1},
\lambda_1)e^{in\lambda_0} \\
F_{\phi}(\lambda_{-1},\lambda_0,\lambda_1)&\equiv&\sum\limits_{n}\phi^n
(\lambda_{-1},\lambda_0,\lambda_1)=\sum\limits_{n}\phi^n
(\lambda_{-1},\lambda_1)e^{in\lambda_0} \nn \\
F_l(\lambda_{-1},\lambda_0,\lambda_1)&\equiv&\sum\limits_{n}l^n(\lambda_{-1},
\lambda_0,\lambda_1)=\lambda_0+\sum\limits_{n\neq 0}l^n(\lambda_{-1},\lambda_1)
e^{in\lambda_0}\;\;\;\;, \nn 
\eea
where the explicit solution to the classical equations of motion has been 
used in the second equality. 

We can take the classical limit $c\rightarrow\infty$, that is, 
$R\rightarrow\infty$ (exactly in the same way as we did in
\cite{qg1199}), after we have made the linear change of variables 
$u=\frac{1}{2}(\lambda_1+
\lambda_{-1}),v=\frac{1}{2}(\lambda_1-\lambda_{-1}),\lambda=\lambda_0$ and 
imposed the Casimir constraint, which compels the previously defined fields 
to {\it live} on $AdS$ spacetime. In this situation we find:
\bea
F_{g_i}(u,\lambda)&\equiv&F_{g^i\;AdS}^{R\rightarrow\infty}(u,\lambda)=
\sum\limits_{n}{g_i}^n(u)e^{in\lambda} \;\;\;,\; i\in\{1,2,3\}  \\
F_{l^n}(u,\lambda)&\equiv&F_{l\;AdS}^{R\rightarrow\infty}(u,\lambda)=
\lambda+\sum\limits_{n\neq 0}{l}^n(u)e^{in\lambda}\;\;. \nn 
\eea
In this limit, it is easy to express the $g_i^n$'s in terms of the $F_{g_i}(u,
\lambda)$ and we can write $\Theta$ in the configuration-like 
variables. By defining the action as $S=\int \Theta$, we encounter:
\bea
S=S^{KM}+S^{Vir}+S^{Int}\;\;\;\;, \nn
\eea
where $S^{KM}$ is the action for pure $SU(1,1)$-Kac-Moody 
\footnote{Here only evaluated in a perturbative manner up to the second order. 
Since we shall not make use of the Lagrangian formalism, we do not give
the explicit expression here. We only notice that at first order the
Kac-Moody modes behave as free scalar fields, which become coupled at 
higher orders.}, $S^{Vir}$ is the corrected Polyakov action for $2D$ quantum 
gravity found in \cite{qg1199}, and $S^{Int}$ is an interaction term, which 
couples the gravitational degrees of freedom coming from the Virasoro algebra
to the new 
dynamical modes of Kac-Moody. The structure of this term is:
\bea
S^{Int}=\int du d\lambda \frac{{\cal F}(F_z,F_{z^*},F_{\phi})\partial_u F_l}
{\partial_{\lambda} F_l}\;\;, 
\eea
where ${\cal F}(F_z,F_{z^*},F_{\phi})$ is a functional of the fields 
$F_z,F_{z^*},F_{\phi}$, whose actual form (obtained in a perturbative
way) is not relevant here.

We finally point out the natural appearance of an interaction term between 
the gravitational
degrees of freedom and the new ones.

\section{Quantum model}
\subsection{Reduction}
In a group approach to quantum theory, the quantum realization of the 
physical
model is accomplished by the construction of an irreducible and unitary 
representation of the chosen physical group. This representation is obtained 
from the regular representation, which is highly reducible, by imposing
certain conditions to the wave functions. These conditions are encoded in
the so-called polarization subalgebra ${\cal P}$, which is a left-invariant
maximal horizontal subalgebra including the 
characteristic subalgebra. That is, it includes ${\cal G}_{\Theta}$ and one 
mode for each conjugated pair of the dynamical degrees of freedom.

Before we proceed to the explicit construction, we redefine the generators
of the algebra
in order to recover exactly the commutators (\ref{algebra}):
\bea
\hat{L}_n&\equiv& i\tilde{X}^R_{l^n}\;,\;\hat{I}\equiv i\Xi\;,\; 
\hat{G}^i_n\equiv\tilde{X}^R_{g^n_i} \label{redef} \\
\hbox{and}\;\;\;\hat{G}^1_n&\equiv&\hat{Z}_n\;,\;
\hat{G}^2_n\equiv\hat{Z^*}_n\;,\;\hat{G}^3_n\equiv\hat{\Phi}_n\;. \nn
\eea

A characteristic feature of infinite-dimensional groups is the possibility
of the appearance of non-equivalent polarizations. Different 
polarizations 
lead to physically different systems, that is, the dynamics are not 
equivalent. In fact, in our case there are two non-equivalent polarizations
due to the presence of the Kac-Moody subgroup. Let us focus ourselves on this 
Kac-Moody 
group and ignore the Virasoro subgroup for the time being. Considering the 
selected ${\cal G}_{\Theta}$ and looking at the 
commutation relations, we find two possibilities:
\bea
{\cal P}^N_{KM} &=& \langle \tilde{X}^L_{\phi^{n\leq 0}},\tilde{X}^L_{z^r}
\rangle  \nn \\
{\cal P}^S_{KM} &=& \langle \tilde{X}^L_{\phi^{n\leq 0}},
\tilde{X}^L_{z^{p\leq 0}},\tilde{X}^L_{{z^*}^{q<0}}\rangle \;\;\;\;.
\eea
The first case, ${\cal P}^N_{KM}$, called {\it natural} polarization, is 
characterised by the presence of all the operators associated with the 
negative (or positive) roots of the semi-simple algebra, whereas in the second 
one,
${\cal P}^S_{KM}$, called {\it standard}, coexist operators associated with all
the roots of the finite algebra (see \cite{KMpepe} for details in 
$SU(2)$-Kac-Moody). In this paper we shall consider only the 
second case, since unitarity compels $\alpha$ to be 
zero in the natural polarization, which is too a 
severe condition and makes the dynamics less interesting in this framework,
although a meaningful gravitational model can still be constructed with this 
natural polarization \cite{Klauder}. 

In a na\"\i ve way, we would expect that the inclusion of the Virasoro modes
would not significatively alter the construction of a {\it standard-like} 
polarization associated with our selected ${\cal G}_{\Theta}$. The union of 
the  
polarization for the two well-studied separate cases (Kac-Moody and Virasoro),
${\cal P}^S={\cal P}^S_{KM}\cup {\cal P}_{\overline{Vir}}$,
seems to be a good candidate for a polarization of the entire group and, in 
fact, this would be the case if the only Virasoro mode in ${\cal G}_{\Theta}$ 
was $\tilde{X}^L_{l^0}$ or if the parameter $K$ was zero. The first 
possibility must be rejected since our aim is to generalize the model in 
\cite{qg1199}, for which the presence of a $sl(2,\mathbb R)$, from the
Virasoro algebra, inside
${\cal G}_{\Theta}$ is fundamental for the spacetime notion. The second one 
must also be discarded on unitarity grounds as will be seen below.

Unfortunately ${\cal P}^S_{KM}$ is not left invariant under the action of 
${\cal P}_{\overline{Vir}}$. Indeed, the commutator $[\tilde{\bar{X}}^
L_{l^1},
\tilde{X}^L_{z^0}]$ is proportional
to $\tilde{X}^L_{z^1}$ (since $K\neq0$), which is absent from the subalgebra
${\cal P}^S_{KM}$.
And what is worse, there is no full polarization
containing the whole characteristic subalgebra \footnote{To be precise, a 
full polarization can be constructed for $\frac{K}{2\alpha}=-1$, but the
resulting representation is not exponentiable, according to the 
{\it fusion rule} to be derived in the next section. The same
globallity requirement excludes the {\it natural-like} full polarizations for
$\alpha\neq 0$.}.
This is an intrinsic, algebraic pathology that cannot be avoided. 

We shall call this situation a $SL(2,\mathbb R)-anomaly$, to distinguish it
from the more usual case in conformal field theories (like WZW models
or string theory) where the entire Virasoro algebra is devoid of dynamical 
content, but nevertheless cannot be included inside the polarization as a 
whole. The latter is conventionally referred to as conformal anomaly.

A well established procedure in the framework of group quantization, whenever
the whole characteristic algebra cannot be included inside the polarization,
is to correct the operators in ${\cal G}_\Theta$ with higher-order terms
in the left enveloping algebra, giving rise to a higher-order characteristic
algebra and/or polarization. 
The simplest physical example exhibiting this solution, either explicit or
implicitly, is the case of the Schr\"odinger group \cite{schr}
in non-linear quantum optics \cite{optica}. The symmetry of this problem is 
that of the harmonic oscillator group ($\hat{a}, \hat{a}^\dagger, \hat{H}$) 
with the two extra generators $\hat{l}_{-1}, \hat{l}_1$, closing with 
$\hat{H}\equiv \hat{l}_0$ a 
$SL(2,\mathbb R)$ algebra, which constitutes the characteristic subalgebra. 
This symmetry is intrisicly anomalous, since no first-order full polarization 
including the $SL(2,\mathbb R)$ can be found. Resorting to the left-enveloping 
algebra, an extra $SL(2,\mathbb R)$ of the form (left version of)
$\frac{(\hat{a})^2}{2}, \hat{a}\hat{a}^\dagger,
\frac{(\hat{a}^\dagger)^2}{2}$ can be found, in such a way that the 
difference with 
the first-order one can be included in the polarization, thus solving the 
anomalous reduction problem.

Following the same reasoning in the present case, we seek new operators in the
left-enveloping algebra. Inspired by the abovementioned example, as well as
the WZW models, the left Sugawara operators in the Kac-Moody-quadratic 
enveloping algebra constitute a natural guess. The standard expressions of 
the left- and right-invariant forms of these generators 
(analogous to $\frac{(\hat{a})^2}{2}, \hat{a}\hat{a}^\dagger,
\frac{(\hat{a}^\dagger)^2}{2}$), in terms of the Non-Pseudo-extended (NP)
generators ${X_{NP}}^{L,R}_{g^n_i}\equiv \tilde{X}^{L,R}_{g^n_i}-\frac{K}{2}
\delta_{n,0}\delta_{i,3}\Xi$, are: 
\bea
(\tilde{X}^{Sug}_{l^n})^{L,R}\equiv\;\frac{1}{2\alpha}:\sum\limits_m k^{ij}
{X_{NP}}^{L,R}_{g^n_i}{X_{NP}}^{L,R}_{g_j^{n-m}}: \;
 \;\;i,j\in \{1,2,3\}\;\;\;\;, \label{Lsugawara}
\eea
where $:\;:$ denotes standard normal ordering and $k^{ij}$ is the Killing 
metricof the rigid group. 
Their commutation relations with the first-order left-invariant vectors are:
\bea
\left[(\tilde{X}^{Sug}_{l^n})^L,\tilde{X}^L_{g_i^m}\right]&=&-im 
\tilde{X}^L_{g_i^{n+m}} \nn\\
\left[(\tilde{X}^{Sug}_{l^n})^L,(\tilde{X}^{Sug}_{l^m})^L\right]&=&
i(n-m)(\tilde{X}^{Sug}_{l^{n+m}})^L+\frac{i}{12} c^{Sug} n^3
\delta_{n+m} \Xi \\
\left[\tilde{X}^L_{l^n},(\tilde{X}^{Sug}_{l^m})^L\right]&=&i(n-m)
(\tilde{X}^{Sug}_{l^{n+m}})^L+\frac{i}{12} c^{Sug} n^3
\delta_{n+m}\Xi\;\;\;\;. \nn
\eea
We see that the Sugawara operators close a Virasoro algebra with  
$c^{Sug}=\frac{\alpha dim(G)}{-\hbox{g}+\alpha}$, $\hbox{g}$ being the dual 
Coxeter 
number
(see \cite{Francesco}). These commutators can 
be classically checked, and the central extensions then fixed by consistency 
with Jacobi identities. Notice that these higher-order operators close
a proper algebra with the first-order ones.

These new generators are used to correct the first-order characteristic algebra
and to define a higher-order polarization in terms of the difference of the 
first- and second-order Virasoro subalgebras: $(\tilde{X}^I_{l^n})^L=
\tilde{X}^L_{l^n}-
(\tilde{X}^{Sug}_{l^n})^L$ \footnote{Strictly speaking, we just need
to correct the $\tilde{X}^L_{l^1}$ generator, but the analysis is simpler if
we extend this correction to the entire Virasoro subalgebra.}. 
The commutation relations for the intrinsic Virasoro generators,
$(\tilde{X}^I_{l^n})^L$, with themselves and the Kac-Moody modes are:
\bea
\left[(\tilde{X}^I_{l^n})^L,(\tilde{X}^I_{l^m})^L\right]&=&
i(n-m)(\tilde{X}^I_{l^{n+m}})^L+\frac{i}{12}[(c-c^{Sug}) n^3-
c'n]\delta_{n+m} \hat{I} \\
\left[(\tilde{X}^I_{l^n})^L,\hat{X}_{g_i^m}\right]&=&0\;\;\;\;, \nn
\eea
and, therefore, the corrected Virasoro group does not display the fatal 
non-diagonal action on the Kac-Moody polarization.
This detail allows us to construct a polarization as the simple union of
the Kac-Moody and the intrinsic Virasoro one, permitting in particular the 
presence of the corrected characteristic subalgebra inside the polarization:
\bea
{\cal P}={\cal P}^S_{KM}\cup\langle(\tilde{X}^I_{l^{n\leq 1}})^L\rangle 
\;\;\;\;.
\eea
The irreducibility of the representation is guaranteed when 
the carrier space is constructed from the orbit of the group 
through a vacuum state $\mid\!0\rangle$. The 
representation becomes a maximum-weight one in which the annihilation
operators are the adjoint\footnote{See next subsection.} versions of the
right-invariant partners of the vector fields in the 
polarization. In parallel to (\ref{redef}) we define: 
$\hat{L}^I_n\equiv i(\tilde{X}^I_{l^n})^R$ and 
$\hat{L}^{Sug}_n\equiv i(\tilde{X}^{Sug}_{l^n})^R$.
Thus, the ultimate outcome of the formal polarization process we have just
described, is to provide the form of the vectors in the representation space:
\bea
\mid\!\Psi\rangle=\prod_{i\in\{1,2,3\}}\hat{G}^i_{n_1}...
\hat{G}^i_{n_i}\hat{L}^I_{p_1}...\hat{L}^I_{p_j}
\mid\!0\rangle \;\;\;(n_i\leq-1 \;if\;i\in \{1,2\},n_3\leq0,
p_{j}\leq-2) \;\;\;\; ,
\eea
where $\mid\! 0\rangle$ is the vacuum state.

\subsection{Unitarity}
Now we shall consider the problem of unitarity. Therefore, we must introduce
a notion of scalar 
product in the representation space. The standard 
way to
implement this in GAQ is by using a measure on the group, which is explicitly 
computed from the group volume. This is a non-trivial issue for finite 
non-compact
groups, although it can eventually be managed. However, in the case of 
infinite-dimensional
groups, as is our case, we encounter a very hard problem. An alternative and
consistent approach is to fix 
the norm of the vacuum state $(\langle 0\!\mid \!0\rangle=1)$ and then choose 
a rule for adjointness of the operators in the representation.

In order to decide the choice of adjoint operators, we impose consistency
with the rigid $SU(1,1)$ subgroup, for which the scalar product
can be introduced in a non-ambiguous way. Concretely, and due to the way
the pseudoextension in the Kac-Moody subgroup has been made, the 
representation 
chosen for the rigid subgroup is associated with the discrete $SU(1,1)\approx
SL(2,\mathbb R)$
series. As can be seen from a explicit construction of this representation
\cite{julio}, we find:
\bea
\hat{Z}^\dagger=-\hat{Z^*}\;,\,
\hat{\Phi}^\dagger=-\hat{\Phi} \;\;\;\;,
\eea
which makes the $SU(1,1)$ representation unitary. These 
relations are translated to the Kac-Moody case in the following way:
\bea
(\hat{Z}_n)^\dagger=-\hat{Z^*}_{-n}\;,\,
(\hat{\Phi}_n)^\dagger=-\hat{\Phi}_{-n} \;\;\;\;.
\eea
This is therefore the rule for adjoint assignment \footnote{For the Virasoro 
modes, as in \cite{qg1199}, we impose $(\hat{L}_n)^\dagger=\hat{L}_{-n}$.} that
fixes our scalar product.

Prior to the unitarity problem  of the representation as regards the scalar
product, another intimately related point must also be considered. Not all 
the algebra representations are lifted to group representations: there are
globallity conditions. In our case, when we restrict ourselves to the 
subgroups generated by $\langle \tilde{X}^L_{z^q},\tilde{X}^L_{{z^*}^q},
\tilde{X}^L_{\phi^0},\frac{\partial}{\partial \varphi}\rangle$, which form 
$SU(1,1)\tilde{\otimes}U(1)$, and consider the polarization given by
$\langle \tilde{X}^L_{z^q},\tilde{X}^L_{{z^*}^q},\tilde{X}^L_{\phi^0}\rangle
\cap{\cal P^S_{KM}}= \langle \tilde{X}^L_{z^q},\tilde{X}^L_{\phi^0}\rangle$,
the global exponentiability of the wave functions forces
the value of the pseudoextension\footnote{Which is directly linked to the 
Bargmann index of the representation.}, $\frac{K}{2}+\alpha n$, to satisfy the 
following inequalities:
\bea
n\leq0\;&,& \;\; \frac{K}{2}+\alpha n\leq-\frac{1}{2}  \\
n>0\;&,& \;\;  \frac{K}{2}+\alpha n\geq \frac{1}{2} \;\;\;\;,\nn
\eea
which imply the $SU(1,1)$-Kac-Moody {\it fusion rule},
\bea
\alpha \geq 0 \;\;\;;\;\;\; -\frac{1}{2}\geq \frac{K}{2}\geq\
-\frac{1}{2}(2\alpha-1)\;. \label{condition}
\eea
Thus, in our scheme, globallity fixes the sign of the central extension 
parameter
$\alpha$ (which would have the opposite sign in the $SU(2)$ case, where the 
inequalities reverse sign) and provides a {\it natural} restriction on the 
permitted values of $K$ according to the actual value of $\alpha$. This second 
condition is analogous to the fusion rule in $SU(2)$ of Kac-Moody. For a 
general analysis of globallity conditions for central extensions of compact 
Kac-Moody groups we refer to \cite{Milikito} (see also \cite{Preisler}).

Now we return to the question of unitarity. If we consider the linear 
combinations:
\bea
\hat{J}^1_n&=&\frac{1}{2}(\hat{Z}_n+\hat{Z^*}_n)\;\;,\;\;
(\hat{J}^1_n)^\dagger=-\hat{J}^1_{-n} \nn \\
\hat{J}^2_n&=&\frac{1}{2}(\hat{Z}_n-\hat{Z^*}_n)\;\;,\;\;
(\hat{J}^2_n)^\dagger=\hat{J}^2_{-n}  \\
\hat{J}^3_n&=&\hat{\Phi}_n \;\;,\;\;(\hat{J}^3_n)^\dagger=
-\hat{J}^3_{-n} \;\;,\nn
\eea
and their commutators:
\bea
\left[\hat{J}^1_n,\hat{J}^1_m\right]&=&-\alpha n \delta_{n+m}\hat{I} \nn \\
\left[\hat{J}^2_n,\hat{J}^2_m\right]&=&\alpha n \delta_{n+m}\hat{I} \\
\left[\hat{J}^3_n,\hat{J}^3_m\right]&=&\alpha n \delta_{n+m}\hat{I}\;, \nn 
\eea
then taking into account the positive sign of $\alpha$, we find that
$\hat{J}^1_n$ and $\hat{J}^2_n$ generate states of positive norm, whereas the 
states generated
by $\hat{J}^3_n$ have a negative one. What we have found is that consistency 
with rigid $SU(1,1)$ representation theory enforces the existence of
negative-norm states in the model, which spoil the unitarity of the theory.
The presence of these states in non-compact Kac-Moody groups is a well-known 
feature (see \cite{Bars,O'Raigh,Peskin,Gawedzki} and references therein). 

According to the standard viewpoint, these states must be eliminated from the 
Hilbert space in order to find the 
{\it physical} quantum phase space. In the case of the bosonic string one 
encounters the same situation, but there the world-sheet reparametrization 
invariance 
compels the (Sugawara) Virasoro modes to act trivially, and this constraint  
eliminates these states. In our case, we cannot resort to the gauge 
invariance  
of a Lagrangian in order to motivate such a constraint. All we can adduce
in our approach is mathematical consistency. In this line, good candidates for
constraints (as indicated in \cite{O'Raigh}) are the Kac-Moody operators
related to the Casimir of the rigid group. In the case of 
$SU(1,1)$, the only such possibility is the set of Virasoro operators.
Therefore, our proposal here is to use the Sugawara-Virasoro operators in 
order 
to eliminate the non-physical states \footnote{A quite different and less
harmful alternative treatment of the unitarity problem in non-compact 
infinite-dimensional groups is under study \cite{fermio}.}. This is a well 
studied problem in the context of a 
bosonic string propagating on a curved spacetime. The answer is that Virasoro 
constraints are not enough to eliminate these vectors in the case of 
$SU(1,1)(\approx SL(2,\mathbb R)$-Kac-Moody (there is not a no-ghost theorem 
in this case).

In \cite{Bars} a number of solutions are proposed and finally discarded.
The first proposed solution is to limit the possible rigid $SL(2,\mathbb R)$
representations present in the Kac-Moody one by imposing the 
index of the $SL(2,\mathbb R)$ (our $K$) to be bounded by the central extension
parameter $\alpha$. This is precisely our second condition in 
(\ref{condition}). This possibility is ruled out in \cite{Bars} because it is
not a natural condition for a string model, since at the quantum level it
eliminates some excitations present in the classical theory. But we are not 
working
with a string theory and therefore there is no reason for excluding this
condition. Even more, this condition is necessary for lifting the 
Kac-Moody algebra to the group level. Thus in our construction, 
is globallity that is responsible for the elimination of negative-norm states.

We can summarize the previous considerations concerning the constraints on the 
Hilbert space that eliminate the negative-norm states \footnote{An important
subtlety of the $\hat{L}^{Sug}_0$ constraint, is that it 
forces an excited Kac-Moody state of level $K$ to be constructed from a vacuum
state with a very concrete non-trivial value of the Casimir of the rigid 
$SU(1,1)$ (see \cite{Bars}). In order to maintain the possibility of different 
excitation levels for the 
matter modes as well as the notion of a true vacuum of the theory, we are 
obliged to consider a direct sum of irreducible representations constructed 
from the true vacuum $\mid\! 0\rangle$ and mixed by {\it external} operators 
which play the same role as the position operators in string theory generating 
translations in the momentum space. We shall not dwell on
any more on this point, since the description of the Kac-Moody Hilbert space to
that degree of accuracy is not crucial for us at this stage.} by writing:
\bea
\hat{L}^{Sug}_n\mid\!\Psi\rangle&=&0\;\;,n\leq 0\;\;\;\; \hat{L}^{Sug}_n
\mid\!\Psi\rangle\sim\mid\! \Psi\rangle\;\;,n>0 \nn \\
-\frac{1}{2}&\geq& \frac{K}{2}\geq -\frac{1}{2}(2\alpha-1) \;\;\;\;,
\eea
where $\sim$ indicates that the two states must be identified by taking the
quotient ($\hat{L}^{Sug}_n\mid\!\Psi\rangle,n>0$ is a spurious state). The
$\hat{L}^{Sug}_0$ constraint, together with the bound on $K$, establish an 
upper limit to the possible excited states that can appear in the theory 
(in fact, this was the reason in \cite{Bars} for rejecting this 
representation).

\subsection{In\"on\"u-Wigner contraction}
There is a mathematical construction devised by In\"on\"u and Wigner 
\cite{IW} which allows the derivation of non-semisimple algebras and their 
representations from the case of semisimple ones by means of a contraction
procedure. For the case of affine Lie algebras this contraction has been 
considered in \cite{majumdar} (see also \cite{montigny}). The effect of this 
contraction can be seen 
as a decrease in the grade of non-linearity of certain contracted modes.
Thus, the dynamics of these modes becomes more linear, which can be 
physically considered as a softening of the associated interaction. This 
mechanism can be of some interest when discussing the classical limit.

Let us briefly recall the fundamentals of the In\"on\"u-Wigner contraction
(for further details see \cite{IW,majumdar}).
For the contraction to be possible, the original algebra must admit a 
decomposition:
\bea
{\cal G}=V_1\oplus V_2 
\eea
such that if $X^\alpha$, $\alpha=1,...,dim V_1$ and $X^i$, $i=1,...,dim V_2$
constitutes a basis for ${\cal G}$, then the Lie-algebra structure constants
${C^{\alpha \beta}}_i$ must vanish.
In this case, we can contract with respect to $V_1$, by redefining the 
generators in
$V_2$ with a multiplicative parameter $\lambda$ that we make tend to zero
($\lambda \rightarrow 0$). After this limit, the $V_1$ subalgebra remains 
unaltered, but the contracted $V_2$ generators undergo a {\it linearization}.
The structure of the resulting algebra is:
\bea
\left[X^\alpha,X^\beta\right]&=&{C^{\alpha \beta}}_\gamma X^\gamma \nn \\
\left[X^\alpha,X^i \right]&=&{C^{\alpha i}}_j X^j  \\
\left[X^i,X^j \right]&=&0 \;\;\;\;.\nn
\eea
In our case, there indeed exists such a $V_1$ subalgebra. It is generated by 
the operators associated with the Cartan rigid subalgebra and the Virasoro 
modes.
Contracting with respect to this subalgebra we find:
\bea
[\hat{Z}_n,\hat{Z^*}_m]&=& 0 \nn \\
\left[ \hat{\Phi}_n,\hat{Z}_m \right] &=& i \hat{Z}_{n+m}\;
,\;
\left[\hat{\Phi}_n,\hat{Z^*}_m\right]=-i \hat{Z^*}_{n+m} \nn \\
\left[\hat{\Phi}_n,\hat{\Phi}_m\right]&=&\alpha n \delta_{n+m} \hat{I} 
\nn \\
\left[\hat{L}_n,\hat{Z}_m\right]&=&-m \hat{Z}_{n+m}\; , \; 
\left[\hat{L}_n,\hat{Z^*}_m\right]=-m \hat{Z^*}_{n+m}
\label{IWalgebra} \\
\left[\hat{L}_n,\hat{\Phi}_m\right]&=&-m \hat{\Phi}_{n+m}+im
\frac{K}{2}\delta_{n+m} \hat{I}\nn \\
\left[\hat{L}_n,\hat{L}_m\right]&=&(n-m)\hat{L}_{n+m}+\frac{1}{12}(cn^3-c'n)
\delta_{n+m}\hat{I}\;\;\;\;. \nn 
\eea
Then we find that, in this limit, the $z$ and $z^*$ modes lose their dynamical
character, and the physical degrees of freedom related to them disappear.
On the other hand, as seen in the previous subsection, the modes associated 
with the Cartan generator are unphysical, so the only physical degrees of 
freedom after the contraction are the Virasoro ones.

It is very important to note that if the Virasoro modes were not present
in the model, a dynamical content could be associated with the contracted modes
as a trace of the pseudoextension $K$. In that case, after the 
In\"on\"u-Wigner contraction,
the Kac-Moody central extension would disappear for this modes, but the 
pseudo-extension would became a real extension (the same phenomenon happens for
the Lorentz and Galileo groups):
\bea
[\hat{Z}_n,\hat{Z^*}_m]&=& \kappa \delta_{n+m}\hat{I} \;\;\;\;. 
\eea
However, the presence of the Virasoro generators in the physical algebra
makes this extension impossible: it is forbidden by the Jacobi identities.
What we have found is that the presence of {\it gravity} modes imposes 
constraints to the size of physical phase space.

We shall comment further on the role of this In\"on\"u-Wigner contraction when
we consider the physical interpretation of the model.

\section{Physical analysis of the model}
In this section we study the physical content of the mathematical
objects presented in the previous sections.

As far as the quantum model is concerned, our chief aim is to give an 
interpretation 
to the states in the Hilbert space. We first focus on the states of the form:
\bea
\hat{L}^I_{n_1}...\hat{L}^I_{n_j}\mid\!0\rangle \;\;\; n_1,...,n_j
\leq-2\;\;\;\;, 
\eea
generated by the intrinsic Virasoro modes. The role of these vectors is that
of generating the underlying spacetime structure, exactly in the same way
the Virasoro modes work in \cite{qg1199} (see this reference for further
details). They span a Virasoro 
irreducible representation with $c^I(=c-c^{Sug})=c'>1$, which is reduced under 
its 
kinematical $sl(2,\mathbb R)$ subalgebra, producing an ensemble of 
$sl(2,\mathbb R)$ irreducible representations, denoted by $R^{(N)}$. An AdS 
spacetime of {\it radius} $R=\frac{c^I}{\sqrt{N(N-1)}}$ is associated with 
each 
$R^{(N)}$ representation, while a specific vector state $\mid\!\!N,n,i\rangle$ 
in $R^{(N)}$ represents a particular state of that spacetime ($n$ is an 
excitation index and $i$ refers to the degeneration of $R^{(N)}$). These 
$\mid\!\!N,n,i\rangle$ states
constitute an orthogonal basis of the Virasoro representation.
The interpretation of the intrinsic Virasoro modes is, therefore, the analogue 
of the Virasoro modes in the pure gravity model, being the basis of the
spacetime skeleton. 

The novel feature is the presence of the {\it matter} degrees of freedom.
A general state can be formally written as:
\bea
\mid\!\Psi\rangle=\sum\limits_{N,n,i} (Physical\; Kac-Moody\; modes)\mid 
\!N,n,i\rangle \;\;\;\;.
\eea
We may think of the states generated by the physical modes $\hat{J}^1_n$ 
and $\hat{J}^2_n$ (or equivalently by $\hat{Z}_n$ 
and $\hat{Z^*}_n$) as being the quanta of some
{\it quantum fields} representing matter. Of course, these are not fields
in the ordinary sense, since we lack of a unique spacetime background 
(we have a whole ensemble of them) on which these objects would have support. 
Only when we consider a state completely lying on a specific spacetime 
(that is, when $N$ and $i$ are fixed in the previous sum), does a standard 
notion
of field show up, and then $\hat{J}^1_n$ and $\hat{J}^2_n$ can be seen as 
{\it excitations}
of them. In general, a physical state is a linear superposition of different
spacetimes, each one supporting a different content of matter excitation 
modes. 
Therefore, matter has an essential global and non-local character.

A very important point is the fact that matter degrees of freedom are not 
free modes, as a consequence of their non-trivial commutation 
relations. Thus, if a state with a given matter field content suffers the 
action of a (matter) perturbation, the 
reordering process (originated when 
passing the lowering operator to the right until it annihilates the 
vacuum) causes a mixing among the matter fields which results in a change in
the distribution of matter modes. This makes the
structure of the Hilbert space a very complicated one, a situation that gets
even worse when the Sugawara constraints (a part of the Virasoro
gravity modes) are taken into account.
Despite the complexity of the quantum phase space, the physical image of the 
system is quite simple: we have two interacting quantum matter fields spread
over different spacetimes.

The effect of gravity is the consequence of the action of the complete
Virasoro modes: $\hat{L}_n=\hat{L}^{Sug}_n+\hat{L}^I_n$. When one of these 
modes acts on a given 
physical state $\mid\!\Psi\rangle$, it has a double effect. On the one hand,
and due to the $\hat{L}^{Sug}_n\;(n>0)$ part, it affects the matter 
distribution (again 
as a consequence of non-trivial commutation) thus creating a 
gravity-matter interaction, and on the other hand, the $\hat{L}^I_n$ changes 
the spacetime distribution of {\it Universe}, in the sense of \cite{qg1199}.
This double effect enriches the dynamics of the model with respect to the
pure gravity case.
We see that the {\it intrinsic} part of Virasoro ($\hat{L}^I_n$) is 
responsible for the
spacetime notion and its dynamics, whereas the {\it orbital} part 
($\hat{L}^{Sug}_n$), absent when
no matter is present, is the responsible of gravity-matter interaction.

The semi-classical limit is accomplished by making $c\rightarrow\infty$. 
In this limit, the different spacetimes collapse into a unique AdS with a very 
large radius. The matter modes are therefore defined on the same 
spacetime support, giving rise to an interpretation as excitations of standard
matter fields. 

The dynamics of these fields can be studied by using the
semiclassical formalism presented in {\bf section 2}. We note that the quantum 
condition for the existence of a spacetime notion, 
$c-c^{Sug}=c'$, becomes indistinguishable from the classical condition $c=c'-
3\frac{K^2}{\alpha}$ in
this limit, since $c^{Sug}$ and $\frac{K^2}{4\alpha}$ are upper-bounded 
quantities, so that both conditions are consistent.
As we saw in detail in that section, a semiclassical action can be constructed
for the classical fields. This action is the sum of a gravity, a matter and
an interaction term. 
The matter term has a perturbative expression which at the lowest order 
represents three free scalar fields. This can be seen directly from the form
of $\Theta^{KM}$ (\ref{teta}). The higher-order correction terms couple these 
scalar fields, producing non-trivial matter dynamics. The non-physical field
$F_\phi$ can be seen as an auxiliary field necessary for implementing the 
matter dynamics.
The gravity term has exactly the same form we found in \cite{qg1199}, and we 
take from there the interpretation for the field $F_l$, as an effective 
classical metric field giving rise to a dynamical correction to the 
kinematical background metric:
\bea
ds^2=ds^2_{AdS(R\gg 1)} \;+\; \partial_\lambda F_l d\lambda du \;. 
\eea
The presence of the gravity-matter interaction term shows the influence of 
matter in the classical metric notion, as desired.

The meaning of the physical matter fields, $F_z$ and $F_{z^*}$ is not quite 
clear. A possibility is that they are simply classical non-free scalar fields 
in interaction with gravity, but this then poses the problem of identifying 
the kind of physical matter or interaction they correspond to. 
But another possibility is that these degrees of freedom only makes sense in
a strong interaction regime, in such a way that in the soft classical limit 
they decouple. The mathematical justification for this option is the
possibility of incorporating an In\"on\"u-Wigner contraction in the model, in
which case the physical matter modes would appear corrected by a 
multiplicative parameter $\lambda$. In the weak interaction limit, 
$\lambda\rightarrow 0$, the modes associated with $z^n$ and ${z^*}^n$ 
lose their dynamical content.
This is a non-trivial fact due to the presence of gravity modes, and 
manifests itself by a decrease in the size of the classical phase space with
respect to the quantum one. We have {\it fields} which possess a physical 
existence only in the non-linear strong regime, and become trivial in the weak 
one. A mechanism like this could be interesting in the study of the softening 
of gravity singularities: it would indicate the existence of {\it exotic} 
physics in the surroundings of singularities which disappear as we move away
from them and with no analogues in the classical theory.

One potential snag in the previous discussion is the specification of the 
nature of the
parameter $\lambda$ . A number of possibilities can be found inside the model
by combining the constants $\alpha, c$ and $c'$. Unfortunately the model is not
sufficiently predictive so as to select a specific one.

\section{Conclusions}
We have tried to formulate the dynamics of gravity in the presence of matter,
assuming the very strong hypothesis that the whole physical content of the 
system is encoded in abstract 
symmetry principles. On this line, we have constructed a quantum model 
whose physical states represent linear superpositions of AdS spacetimes 
with different radii and where matter excitations have a definite non-local
character; in the general case, they have support on various 
spacetimes simultaneously. However, a notion for the probability of a matter 
excitation to lie on a specific spacetime makes sense due to the orthogonality
of spacetime vectors and the trivial commutation of matter and intrinsic
gravity modes. It should be remarked that the quantum realization of spacetime 
is directly associated with the $SL(2,\mathbb R)$-anomaly,
eventually solved by means of a higher-order polarization closing 
a proper algebra. (In general, higher-order polarizations only have to close 
weakly, i.e. on solutions.) This allows a well-defined integral manifold
supporting the reduced wave functions.

Global features of the symmetry structure have proven to be crucial for the 
consistency of the model. They have allowed us to eliminate the {\it ad hoc} 
character of the crucial restrictions 
(\ref{condition}), by deducing the natural and unavoidable necessity of them
\footnote{This strengthens the relationship between unitarity and globallity.}.
In particular, these restrictions have endowed the matter degrees of freedom 
with a peculiar behaviour in the quantum regime, since it limits the excitation
capabilities of these modes. 

We have found the phenomenon of an enlarging of the physical phase 
space at the
quantum level; that is, the quantum emergence of physical degrees of 
freedom which are absent from the classical limit. This can be seen both in  
the quantum acquisition of dynamical content of the diffeomorphisms and in
the classical loss of physical presence of the matter modes via an
In\"on\"u-Wigner contraction process. This is an intrinsically interesting 
question which probably survives this particular model and could be present in 
completely different physical systems.

Finally, we recognise the essential limitations and difficulties of 
taking the symmetry hypothesis to its ultimate consequences in the concrete
formulation of this model. Even though it has 
an intrinsic beauty and power, it poses ambiguity problems related
to the physical interpretation of the constructed objects. It does not seem to
be strong enough to fix a unique possibility among the
different choices it raises. 
However, we argue that this approach 
provides profound physical
insight when inserted in a more general framework. 
In the context of gravity, the inclusion of genuinely metric notions should
enrich the physical system by suggesting new solutions, so the extension 
of this approach to higher dimensions is an urgent necessity.

\section{Acknowledgements}
We thank J. Guerrero for very useful and illuminating discussions.

\end{document}